\documentstyle[12pt,aasms4,psfig]{article}

\newcommand\pp     {$\pm$}
\newcommand\pers     {s$^{-1}$}
\newcommand\micros  {$\mu$s}

\righthead{QPO phase lags in XTE J1550--564}
\slugcomment{Accepted for publication in ApJ Letters, 29 September 1999}

\begin{document}

\title{The Complex Phase Lag Behavior of the 3--12 Hz Quasi-Periodic
Oscillations during the Very High State of XTE J1550--564}

\author{Rudy Wijnands\footnote{Chandra Fellow. Present address: Center
for Space Research, MIT, Cambridge, MA 02139, USA}, Jeroen Homan, \&
Michiel van der Klis}

\affil{Astronomical Institute ``Anton Pannekoek'', University of
Amsterdam, and Center for High Energy Astrophysics, Kruislaan 403,
NL-1098 SJ Amsterdam, The Netherlands; rudy@astro.uva.nl,
homan@astro.uva.nl, michiel@astro.uva.nl}

\begin{abstract}
We present a study of the complex phase lag behavior of the
low-frequency ($<$20 Hz) quasi-periodic oscillations (QPOs) in the
X-ray transient and black-hole candidate XTE J1550--564 during its
very high state. We distinguish two different types of low-frequency
QPOs, based on their coherence and harmonic content. The first type is
characterized by a 6 Hz QPO with a Q (the QPO frequency divided by the
QPO width) of $< 3$ and with a harmonic at 12 Hz. The second type of
QPO is characterized by a 6 Hz QPO with a Q value of $> 6$ and with
harmonics at 3, 12, 18, and possibly at 9 Hz. Not only the Q values
and the harmonic content of the two types are different, but also
their phase lag behavior. For the first type of QPO, the low energy
photons ($<$5 keV) of both the 6 Hz QPO and its harmonic at 12 Hz lag
the hard energy photons ($>$5 keV) by as much as 1.3 radian.  The
phase lags of the second type of QPO are more complex. The soft
photons ($<$ 5 keV) of the 3 and the 12 Hz QPOs {\it lag} the hard
photons ($>$ 5 keV) by as much as 1.0 radian. However, the soft
photons of the 6 Hz QPO {\it precede} the hard ones by as much as 0.6
radian. This means that different harmonics of this type of QPO have
different signs for their phase lags. This unusual behavior is hard to
explain when the lags are due to light-travel-time differences between
the photons at different energies, e.g., in a Comptonizing region
surrounding the area where the QPOs are formed.

\end{abstract}

\keywords{accretion, accretion disks --- stars: individual (XTE
J1550--564) --- stars: neutron --- X-rays: stars}

\section{Introduction \label{intro}}

The soft X-ray transient and black-hole candidate XTE J1550--564 was
discovered early September 1998 (Smith 1998) with the All Sky Monitor
on board the {\it Rossi X-ray Timing Explorer} ({\it
RXTE}). Subsequent observations with the {\it RXTE} Proportional
Counter Array during the initial rise, showed 0.08--4 Hz
quasi-periodic oscillations (QPOs) in the power spectrum, superimposed
on strong ($\sim$30\% rms amplitude) band-limited noise (Cui et
al. 1999). On 19--20 September 1998, a strong X-ray flare was detected
with a peak luminosity of $\sim$6.5 Crab (Remillard et
al. 1998). During this flare, QPOs near 185 Hz were discovered
(McClintock et al. 1998; Remillard et al. 1999) simultaneously with
low-frequency ($<$20 Hz) QPOs, strongly indicating that XTE J1550--564
was in the very high state (VHS) during these observations.  The X-ray
spectral properties and the rapid X-ray variability during the first
part of the outburst were discussed by Cui et al. (1998), Sobczak et
al. (1999), and Remillard et al. (1999).  A second VHS episode began
on March 4 1999, when QPOs near 280 and 6 Hz were detected (Homan,
Wijnands, \& van der Klis 1999a).  These QPOs and the other rapid
X-ray variability will be discussed by Homan et al.  (1999b).  In this
{\it Letter}, we concentrate on the small subset of the observations
discussed by Homan et al. (1999b) when XTE J1550--564 was in the VHS,
showing QPOs above 100 Hz and around 6 Hz. We report on the very
complex phase lag behavior of the low-frequency QPOs.

\section{Observation and selection method \label{observations}}

Our primary goal is to study the phase lag behavior of the
low-frequency QPOs when XTE J1550--564 was in the 1999 VHS episode
(see Homan et al.~1999b).  The public {\it RXTE} observations used are
listed in Table~\ref{tab:obslog}.  Data were accumulated in several
different observational modes, which were simultaneously active.  We
used the 'Binned' mode data, with a time resolution of 4 ms in 8
photon energy bands (covering 2--13.1 keV), and the 'Event' mode data,
with a time resolution of 16 \micros~in 16 bands (13.1--60 keV).  To
determine the QPO properties, we calculated power and cross spectra
using 16 or 256-s intervals and a Nyquist frequency of 128 Hz. To
correct for the dead-time effects on the lags, we subtracted the
average 50--125 Hz cross-vector from the cross spectra (see van der
Klis et al. 1987).

\begin{figure}[]
\begin{center}
\begin{tabular}{c}
\psfig{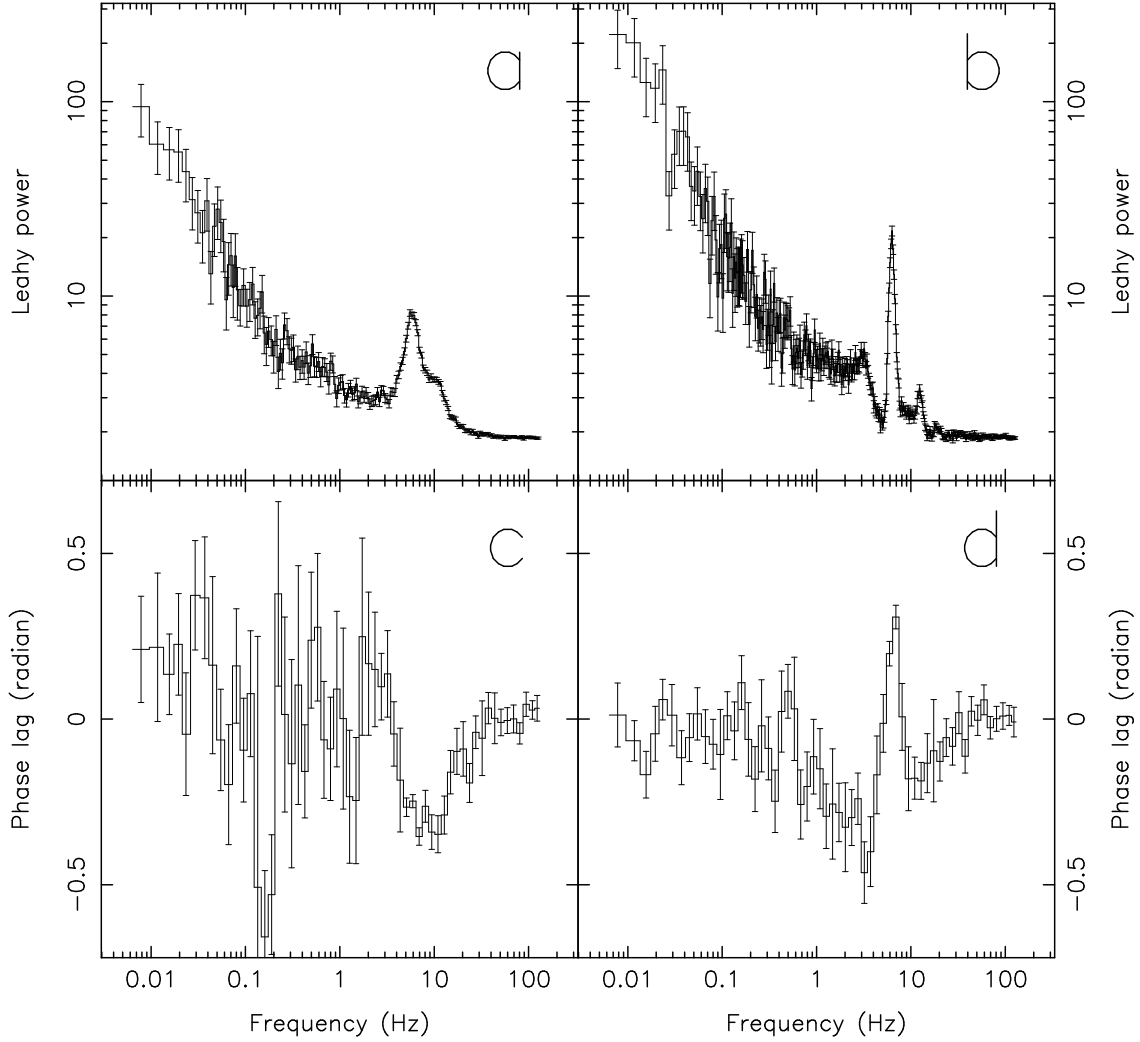}
\end{tabular}
\figcaption{Typical Leahy normalized power spectra (2--48 keV) and
phase lag spectra of observations with type A low-frequency QPOs
({\it a} and {\it c}; observation 40401-01-50-00) and of those with
type B low-frequency QPOs ({\it b} and {\it d}; observation
40401-01-53-00).  In {\it a} and {\it b}, the dead-time modified
Poisson level has not been subtracted.  The phase lag spectra ({\it c}
and {\it d}) were calculated between the energy bands 2.5--6.5 and
6.5--48.0 keV. Negative phase lags mean that the soft photons lag the
hard photons.
\label{fig:powerspectra} }
\end{center}
\end{figure}

The VHS observations can be divided according to the properties of the
QPOs below 20 Hz.  We distinguish two types based on the Q value
(frequency divided by FWHM) of the 6 Hz QPO\footnote{Although the
frequency of this QPO varied between 5 and 7 Hz, for reasons of
clarity this QPO will be referred to as the '6 Hz QPO'.}, and its
harmonic structure: one type with a relatively broad (Q $< 3$) 6 Hz
QPO with a harmonic at 12 Hz (type A QPOs; \S~\ref{section:groupone};
Fig.~\ref{fig:powerspectra}{\it a}), and no other detectable
harmonics, and one with a relatively narrow (Q $> 6$) 6 Hz QPO with
harmonics at 3, 12, 18 (although not always detectable due to the
limited statistics), and possibly at 9 Hz (type B QPOs;
\S~\ref{section:grouptwo}; Fig.~\ref{fig:powerspectra}{\it b}).  The
observations are listed in Table~\ref{tab:obslog} according to the
type of QPO they contain.  Although the power spectra of the
40401-01-57-00 and 40401-01-58-01 observations show fewer harmonics
and stronger band-limited noise, these observations are classified as
containing type B QPOs because of the high Q ($> 6$) values of the 6
Hz QPOs.

\section{Type A low-frequency QPOs
\label{section:groupone}}

Figure~\ref{fig:powerspectra}{\it a} shows a typical power spectrum
(i.e., for observation 40401-01-50-00) containing type A QPOs. Clearly
visible is the 6 Hz QPO with a shoulder at higher frequencies. When
using the data combined for the total {\it RXTE} energy range
(effectively 2--60 keV) and fitting the data with two Lorentzian or
two Gaussian functions, the frequency of the second QPO is not twice
the frequency of the 6 Hz QPO. The ratio between the frequency of the
second QPO and the frequency of the 6 Hz QPO is 1.5--1.8 (depending on
observation and on the fit function [Lorentzian or Gaussian]).
Fitting the total energy band power spectrum with two QPO functions
which are harmonically related to each other, results in fits which
are unacceptable.  However, by using only the data in the range
13.1--48 keV (or 8.0--48 keV when the statistics did not allow to
detect the 12 Hz QPO in the 13.1--48 keV range), the 6 Hz QPO is only
marginally detectable and the most~pronounced QPO is the 12 Hz
QPO. The ratio between its frequency and that of the 6 Hz QPO
(1.91--2.06; depending on observation) indicates that a harmonic
relation with the 6 Hz QPO is likely.  Clearly, the structure of the
power spectrum is more complex than two harmonically related QPOs.
Either the QPO shapes are more complex than Lorentzian or Gaussian
functions, or an extra noise component (maybe a QPO) is present
between the two QPOs.

A typical phase lag spectrum (i.e., for observation 40401-01-50-00)
calculated between 2.5--6.5 and 6.5--48.0 keV is shown in
Figure~\ref{fig:powerspectra}{\it c}.  It can clearly be seen that in
the frequency range 6--12 Hz, the soft photons ($<$ 6.5 keV) lag the
hard ones ($>$ 6.5 keV) by $\sim$0.3 radian. Negative lags mean that
the soft photons arrive later than the hard ones (a soft lag).  The
phase lag of the power law noise component below 1 Hz was consistent
with being zero (0.03\pp0.04 radian for the frequency range 0.01--1
Hz). An extrapolation of the phase lags (assuming a constant phase
lag) of this noise component into the QPO frequency range cannot
explain the phase lags observed for the QPOs. The lags determined for
the QPOs must therefore be intrinsic to the QPOs.  In the phase lag
spectrum, the QPOs cannot be distinguished from each other or from the
possible extra noise component in between them.  Therefore, any
interpretation of the QPO lags should be performed with caution.

In order to determine the phase lags as a function of photon energy,
the frequency and the FWHM of both QPOs are needed.  We decided to
determine the frequency and FWHM of the 6 Hz QPO by fitting a Gaussian
function in the total energy band. A Gaussian function fitted slightly
better than a Lorentzian, although both functions yielded acceptable
fits. Because the type B QPOs need to be fitted with Gaussian
functions in order to obtain acceptable fits
(\S~\ref{section:grouptwo}), we decided, for consistency, to use also
Gaussian functions for the type A QPOs (note that a Gaussian function
will result in a somewhat smaller FWHM than a Lorentzian
function). After determining the properties of the 6 Hz QPO (FWHM 2--5
Hz; frequency 5.2--6.0 Hz) by using the full energy range, we
determined the FWHM (2.5--10 Hz) of the 12 Hz QPO in the energy range
above 8.0 or 13.1 keV by fixing its frequency to twice the 6 Hz QPO
frequency. Usually, the fits were acceptable, although broad excess
noise was sometimes present under the 12 Hz QPO. When including an
extra component in the fit to account for this noise, the FWHM of the
12 Hz QPO was not significantly altered.

Using the thus obtained QPO parameters, we determined the frequencies
and the FWHM of the QPOs for the different type A observations. We
determined the phase lags between different energy ranges by
calculating the average lags in the frequency range determined by the
QPO FWHM centered on the QPO frequency. As a reference band, we used
the 4.4--5.1 keV band.  Figure~\ref{fig:phaselags}{\it a} shows the
phase lags as a function of photon energy for observation
40401-01-50-00.  The soft photons ($<$5 keV) of the 6 Hz QPO ({\it
open squares}) lag the hard ones ($>$5 keV) by $\sim$1.3 radian; the
soft photons of the 12 Hz QPO ({\it open triangles}) lag the hard ones
by $\sim$ 0.6 radian (they are $<3\sigma$ different from the 6 Hz QPO
lags).  The 6 Hz QPO lags in all the other observations with type A
QPOs were consistent with these lags. Due to limited statistics, the
12 Hz QPO lags could only significantly be determined for observations
40401-01-50-00 and 40401-01-51-00. The measured lags and the upper
limits obtained for the other observations were consistent with each
other.

\section{Type B low-frequency QPOs
\label{section:grouptwo}}

Figure~\ref{fig:powerspectra}{\it b} shows a typical power spectrum
(i.e, for observation 40401-01-53-00) containing type B low-frequency
QPOs. Clearly visible is the very significant ($>50\sigma$) 6 Hz QPO
(with Q $ > 6$) and those near 3, 12, and 18 Hz. The 3 Hz QPO seems to
be the fundamental. Between the 6 and 12 Hz QPOs excess noise near 9
Hz is present, possibly due to another harmonic.  A typical phase lag
spectrum is shown in Figure~\ref{fig:powerspectra}{\it d} (calculated
between 2.5--6.5 and 6.5--48.0 keV). Owing to the fact that the QPOs
are relatively narrow, the individual components in the phase lag
spectrum can more easily be distinguished than for the type A QPOs.
Most striking is the fact that the 3 and 12 Hz QPO have {\it soft}
lags (meaning that the photons below 6.5 keV arrive later than the
ones above 6.5 keV) of 0.3--0.4 radian, whereas the 6 Hz QPO has a
{\it hard} lag of 0.3 radian. Thus, the lags in the different QPO
harmonics have different signs.

The power law noise between 0.01--0.5 Hz had a marginally significant
soft lag of 0.05\pp0.02 radian. Above 0.5 Hz the lag increased (and
became significant) to about 0.23\pp0.03 radian between 1 and 2.5
Hz. This is the frequency range where an extra noise component between
1 and 3 Hz is present, indicating that the observed lags in this
frequency range are most likely those for this extra noise component
and not for the power law noise. The above reported lag for the 3 Hz
QPO could also be (partly) due to the lag of this extra noise
component, but it is difficult to disentangle the two
components. Clearly, an extrapolation of the soft lags (assuming a
constant phase lag) below 3 Hz into the 6 Hz QPO frequency range
cannot explain the hard lags observed for this QPO.

To determine the frequencies and the FWHM of these QPOs, we fitted the
power spectra with several Gaussian functions (one for each QPO) that
were harmonically related (using Lorentzians functions resulted in
unacceptable fits).  By including a power law function for the noise
component below 1 Hz, and two extra Gaussians to fit the excess noise
at 9 Hz and the noise between 1 and 3 Hz, we obtained acceptable fits
with $\chi^2_{\rm red} \sim 1$.  In this way, we obtained the
frequencies and FWHMs of the various QPO harmonics.  We determined the
phase lags of the 3, 6, and 12 Hz QPOs in different energy ranges by
calculating the lags in a frequency range determined by the QPO FWHM
centered on the QPO frequency.  As a reference band we again used the
4.4--5.1 keV band. The statistics for the 18 Hz QPOs were not
sufficient to allow detections of its lags.  During some observations
also the statistics of the other QPOs were not sufficient to allow
significant detections.

\begin{figure}[]
\begin{center}
\begin{tabular}{c}
\psfig{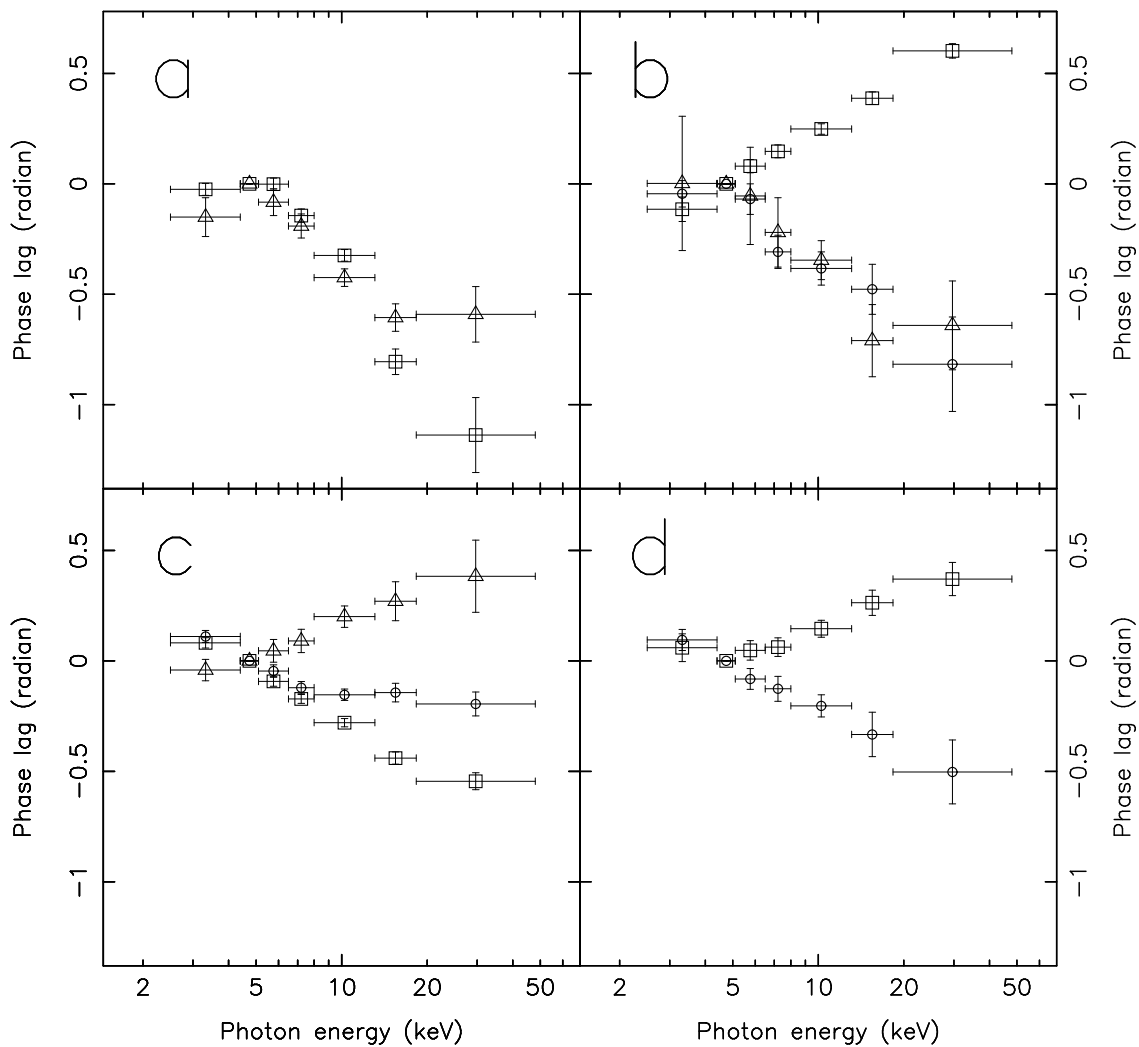}
\end{tabular}
\figcaption{The phase lags of the 3 Hz ({\it open circles}), 6
Hz ({\it open squares}), and 12 Hz ({\it open triangles}) QPOs versus
the photon energy of observation 40401-01-50-00 ({\it a}),
40401-01-53-00 ({\it b}), 40401-01-57-00 ({\it c}), and 40401-01-58-01
({\it d}). The phase lags were calculated with respect to the
reference energy band of 4.4--5.1 keV.
\label{fig:phaselags} }
\end{center}
\end{figure}

A typical example of the resulting phase lags (i.e., for observation
40401-01-53-00) as a function of photon energy is shown in
Figure~\ref{fig:phaselags}{\it b}.  The soft photons ($<$5 keV) of the
3 and 12 Hz QPOs {\it lag} the hard ones ($>$5 keV) by as much as 0.6
radian; the soft photons of the 6 Hz QPO {\it precede} the hard ones
by about 0.6 radian. Again it is clearly visible that the phase lags
of the different QPO harmonics have different signs.  The QPO lags in
the other observations with type B QPOs (except for observations
40401--01--57--00 and 40401--01--58--01, see below) were consistent
with these lags, although for observation 40401--01--51--01 the hard
lags of the 6 Hz QPOs were only 0.3 radian, i.e., half what is
observed for the other observations (the soft lags in the 3 and 12 Hz
QPOs during this observation are consistent with those obtained for
the QPOs during the other observations).

\begin{figure}[]
\begin{center}
\begin{tabular}{c}
\psfig{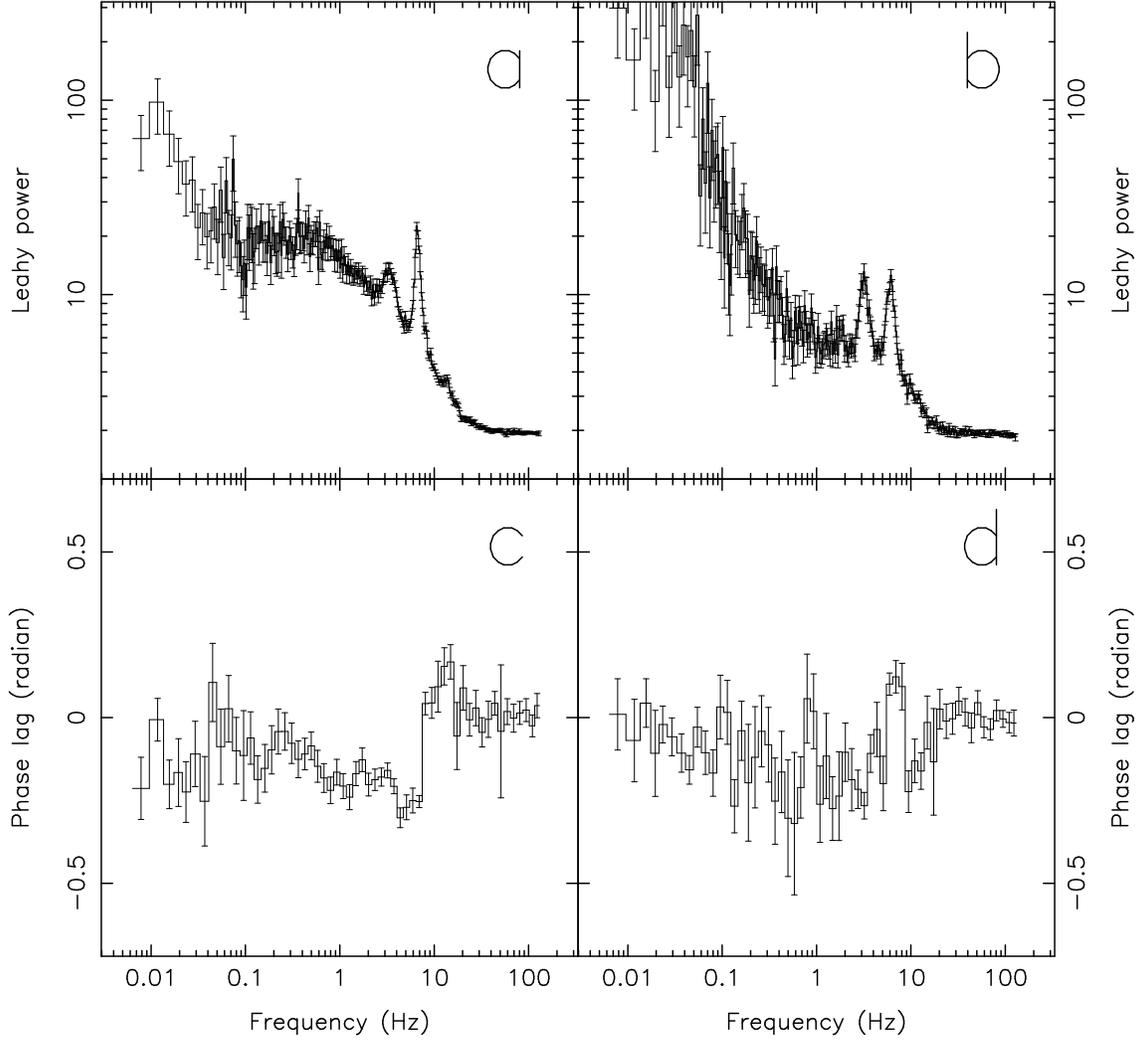}
\end{tabular}
\figcaption{Leahy normalized power spectra (2--48 keV) of the
observations 40401-01-57-00 ({\it a}) and 40401-01-58-01 ({\it b}),
with their corresponding phase lag spectra ({\it c} and {\it d},
respectively).  In {\it a} and {\it b}, the dead-time modified Poisson
level has not been subtracted. The phase lag spectra ({\it c} and {\it
d}) were calculated between the energy bands 2.5--6.5 and 6.5--48.0
keV. Negative phase lags mean that the soft photons lag the hard
photons.
\label{fig:powerspectra2} }
\end{center}
\end{figure}

The QPOs during the 40401--01--57--00 and 40401--01--58--01
observations do not follow the general type B picture.  Their power
spectra are shown in Figures~\ref{fig:powerspectra2}{\it a} and {\it
b}. During observation 40401--01--57--00, a strong band-limited noise
is present, besides the type B QPOs (the Q is $\sim$6 and QPOs at 3
and 12 Hz are visible).  The corresponding phase lag spectrum is shown
in Figure~\ref{fig:powerspectra2}{\it c}. Up to about 8 Hz the soft
photons (below 6.5 keV) lag the hard ones (above 6.5 keV). Around 8--9
Hz a sudden jump in sign occurs and the lags become positive.  The
measured lags for these QPOs are likely a combination of the intrinsic
QPO lags and the noise lags. The 3 Hz QPO phase lag is heavily
affected by the lags of the underlying noise component; the 6 Hz QPO
is also affected, though less than the 3 Hz QPO.
Figure~\ref{fig:phaselags}{\it c} shows the obtained phase lags versus
photon energy for the different QPOs.  Clearly, the lags measured in
the frequency range of 3 and 6 Hz QPOs now have the same sign: the
soft photons {\it lag} the hard ones by 0.2 and 0.3 radian,
respectively. However, the 12 Hz QPO now has a reversed sign: the soft
photons {\it precede} the hard ones by $\sim$0.2 radian. This behavior
is quite different from what is observed for the other type B QPOs.

During observation 40401--01--58--01, the source probably made a state
transition within several minutes (see Homan et al. 1999b). We only
used the data after this transition when two QPOs at 3 and 6 Hz were
visible in the power spectrum (see Fig.~\ref{fig:powerspectra2}{\it
b}). An extra noise component in the same frequency range as the QPOs
is visible.  The phase lag spectrum of this observation is shown in
Figure~\ref{fig:powerspectra2}{\it d}. It is impossible to distinguish
the 3 Hz QPO lag from that of the noise component. The combination of
these two components results in a soft lag of $\sim 0.2$ radian. The 6
Hz QPO has hard lags. The absolute amplitude of its lag ($\sim 0.2$
radian) is smaller than those of the other type B QPOs. This is most
likely due to dilution by the broad band-limited noise component,
which has soft lags. The phase lags of the two QPOs versus energy is
shown in Figure~\ref{fig:phaselags}{\it d}.

\section{Discussion \label{discussion}}

We have presented the complex phase lag behavior of the low-frequency
QPOs which are observed during the March 1999 VHS episode of XTE
J1550--564. We distinguish two QPO types: one type with a relatively
broad (Q $< 3$) 6 Hz QPO with a harmonic at 12 Hz, and one with a
relatively narrow (Q $> 6$) 6 Hz QPO with harmonics at 3, 12, 18, and
possibly at 9 Hz.  The QPO phase lag behavior is different for both
types. The first type always has soft lags, with a maximum amplitude
of 0.6--1.3 radian. For the other type, the different harmonics have
lags of different sign. The absolute maximum amplitude of the lags are
also 0.6--1.3 radian. These phase lags are intrinsic to the
QPOs. Extrapolation of the phase lags (assuming a constant phase lag)
of the power law noise at frequencies lower than the QPO frequencies
cannot explain the observed QPO lags. The noise lags are consistent
with zero, with upper limits of 0.06--0.15 radian, which is
significantly lower than the QPO lags (0.6--1.3 radian).  When an
extra noise component (in addition to the power law noise) is present
in the QPO frequency range, the phase lag behavior becomes more
complex, but detailed studies of the lags are difficult due to the
dilution by this noise component.

Our results show that the phase lag behavior of the low-frequency QPOs
in XTE J1550--564 is quite complex. Especially the different signs for
the lags of different harmonics are unexpected.  This different sign
strongly indicates that differences in light-travel time between the
soft and hard photons cannot account for the lags.  In such a
situation, one would expect that the lags have the same sign for all
the harmonics. This rules out the possibility that the lags are
entirely due to a Comptonizing region surrounding the area where the
QPOs are produced.  The different signs demonstrate that the QPO
waveform is significantly different at low photon energies with
respect to that at higher energies.  The same mechanism which produces
the QPOs most likely also causes this difference.

Only for the black-hole transient GS 1124--683 the phase lags for
low-frequency VHS QPOs have been measured before (Takizawa et
al. 1997). Because we see significant differences in the results in
XTE J1550--564 between different observations, it is difficult to
compare the results between these two sources. The power spectrum of
the 11 Jan 1991 observation of GS 1124--683 (see Fig.~3 of Takizawa et
al. 1997) shows similar QPOs as in our power spectra with type B QPOs:
several harmonically related QPOs with a Q of $\sim$6 for the 6 Hz
QPO.  The lags for the 6 Hz QPO in GS 1124--683 (between photons above
3 keV and those at 3 keV) have similar sign (hard lags) but smaller
amplitudes (0.2--0.4 radian) as our type B QPOs. Another possible
difference is that in GS 1124--683 also the photons below 3 keV lag
the $\sim$3 keV ones. This energy range cannot accurately be probed
with {\it RXTE}, and similar behavior cannot be excluded in XTE
J1550--564. The non-detection of any lags for the noise is consistent
with our results.  The main difference is that for GS 1124--683 the
lags have the same sign for both the 3 as the 6 Hz QPO (see Takizawa
et al. 1997), while for the type B QPOs for XTE J1550--564 the signs
are different. Also, during our only observation with strong
band-limited noise (40401-01-57-00), the noise {\it and} the QPOs had
soft lags, contrary to what is seen for GS1124--683. A uniform picture
for the low-frequency black-hole QPOs cannot be constructed based on
the results so far available.

The only other systems for which the phase lags for the low-frequency
QPOs have been studied are the neutron star low-mass X-ray binaries
(LMXBs). QPOs around 4--7 Hz are observed in the intrinsically
brightest of these systems (van der Klis 1995), and occasionally in
the lower-luminosity ones (Wijnands, van der Klis, \& Rijkhorst 1998;
Wijnands \& van der Klis 1999; Revnivtsev et al. 1999). However, so
far, no harmonics have been observed and the phase lags of these QPOs
are either much larger (up to $\pi$ radian; Mitsuda \& Dotani 1989) or
consistent with zero (e.g., Wijnands et al. 1999).  The other type of
low-frequency QPO seen in the neutron star LMXBs, are the 15--70 Hz
QPOs.  Often, two QPOs are seen which are harmonically related. Their
lags are around 0.3--0.6 radian, but the hard photons always lag the
soft ones (e.g., Vaughan et al. 1994), which is different from XTE
J1550--564.  Also, in these neutron star QPOs the amplitude of the
phase lags of the fundamental is about half that of its second
harmonic. In XTE J1550--564, the absolute phase lags for the different
harmonics are very similar. On the basis of the lags it is doubtful if
the QPOs in the neutron star LMXBs are related to the low-frequency
QPOs in XTE J1550--564.

\acknowledgments

This work was supported by ASTRON (grant 781-76-017), by NOVA, and by
NASA through Chandra Postdoctoral Fellowship grant number PF9-10010
awarded by CXC, which is operated by SAO for NASA under contract
NAS8-39073.  This research has made use of data obtained through the
HEASARC Online Service, provided by the NASA/GSFC.

\clearpage

\begin{deluxetable}{lllll}
\tablecolumns{5}
\tablewidth{0pt}
\tablecaption{A log of the observations\label{tab:obslog}}
\tablehead{
Type of QPO & Obs. ID        & Time                   & Good Time
&Average count rate$^a$\\
      &                &  (March 1999; UTC)     &  (ksec) &  counts
\pers~($\times 10^3$) \\}
\startdata
A     & 40401-01-50-00 &  4 19:14--20:43        & 3.1 & 20.9\\
      & 40401-01-51-00 &  5 12:10--13:01        & 1.5 & 20.1\\
      & 40401-01-59-01 & 18 02:04--02:29        & 1.1 & 11.9\\
      & 40401-01-59-00 & 18 03:38--05:40        & 4.0 & 11.7\\
      & 40401-01-61-01 & 21 02:07--02:27        & 0.9 & 9.5\\
      & 40401-01-61-00 & 21 11:55--12:20        & 1.1 & 9.3\\
\hline	
B     & 40401-01-53-00 &  8 08:29--09:17        & 2.7 & 22.9\\
      & 40401-01-55-00 & 10 23:27--00:31        & 2.8 & 21.9\\
      & 40401-01-51-01 & 11 02:12--02:36        & 1.3 & 22.0\\
      & 40401-01-56-00 & 12 09:36--09:58        & 0.8 & 20.4\\
      & 40401-01-56-01 & 12 11:17--11:26        & 0.4 & 20.6\\
      & 40401-01-57-00 & 13 16:29--17:36        & 2.9 & 12.8\\
      & 40401-01-58-00 & 15 05:15--05:55        & 1.2 & 17.6\\
      & 40401-01-58-01 & 17 02:03--03:16        & 1.4 & 16.5\\
\enddata 
\tablenotetext{a}{The count rates are for 5 detectors on, and are background
subtracted, but not dead-time corrected.}
\end{deluxetable}

\end{document}